\documentclass[conference]{IEEEtran}
\usepackage{graphicx,amsmath,amssymb,cite,algorithm,algorithmic}

\begin{document}
\title{A Reconfigurable Distributed Algorithm\\ for $K$-User MIMO Interference Networks}
\author{
\authorblockN{George~C.~Alexandropoulos and Constantinos B. Papadias}
\authorblockA{Broadband Wireless and Sensor Networks Group, Athens Information Technology (AIT)\\ 19.5 km Markopoulo Avenue, 19002 Peania, Athens, Greece}
e-mails: \{alexandg, cpap\}@ait.gr}
\maketitle

\begin{abstract}
It is already well-known that interference alignment (IA) achieves the sum capacity of the $K$-user interference channel at the high interference regime. On the other hand, it is intuitively clear that when the interference levels are very low, a sum-rate scaling of $K$ (as opposed to $K/2$ for IA) should be accessed at high signal-to-noise ratio values by simple (``myopic'') single-link multiple-input multiple-output (MIMO) techniques such as waterfilling. Recent results have indicated that in certain low-to-moderate interference cases, treating interference as noise may in fact be preferable. In this paper, we present a distributed iterative algorithm for $K$-user MIMO interference networks which attempts to adjust itself to the interference regime at hand, in the above sense, as well as to the channel conditions. The proposed algorithm combines the system-wide mean squared error minimization with the waterfilling solution to adjust to the interference levels and channel conditions and maximize accordingly each user's transmission rate. Sum-rate computer simulations for the proposed algorithm over Ricean fading channels show that, in the interference-limited regime, the proposed algorithm reconfigures itself in order to achieve the IA scaling whereas, in the low-to-moderate interference regime, it leads itself towards interference-myopic MIMO transmissions. 
\end{abstract}
\thispagestyle{empty}
\IEEEpeerreviewmaketitle

\section{Introduction}\label{sec:Intro}
Interference alignment (IA) is a recently proposed transmission technique for the $K$-user interference channel which is shown to achieve a sum-rate multiplexing gain of $K/2$ at the high interference regime \cite{C:Maddah-Ali, J:Jafar_interference}. IA is based on appropriate linear precoding at the transmitters, aiming at post-receiver processing interference cancellation, and requires only global channel state information at all participating transceivers.

Exploiting the space dimension of multiple-input multiple-output (MIMO) systems to perform IA, several research works presented IA-achieving precoding designs \cite{C:Maddah-Ali, J:Jafar_interference, C:Schmidt2009, C:Heath_Globecom, J:Iterative_Jafar, J:Sung_TWC2010, J:Heath_cooperative, C:NegroSlock2011, C:Hadi, J:Luo_TIT_2012} and investi\-gated the feasibility of IA \cite{J:Yetis, J:Gou_KMNchannels} for the $K$-user MIMO interference channel. For the special case of $K=3$, a closed-form solution for IA was presented in \cite{J:Jafar_interference} that was further processed in \cite{J:Sung_TWC2010} for increased sum-rate performance. However, for $K>3$ MIMO communicating pairs, closed-form solutions for IA are in general unknown and several iterative algorithms have been recently proposed (see e$.$g$.$ \cite{C:Schmidt2009, C:Heath_Globecom, J:Iterative_Jafar, J:Heath_cooperative, J:Sung_TWC2010, C:NegroSlock2011, C:Hadi, J:Luo_TIT_2012} and references therein). The vast majority of those algorithms targets at implicitly achieving IA through the optimization of a constrained objective function. To this end, several objective functions have been considered, such as for example: \textit{i}) minimization of the total interference leakage \cite{J:Iterative_Jafar, C:Hadi}, \textit{ii}) minimization of the sum of squared errors \cite{C:Heath_Minimization}, \textit{iii}) minimization of the mean squared error (MSE) \cite{C:Schmidt2009, J:Heath_cooperative, J:Luo_TIT_2012}, \textit{iv}) maximization of the signal-to-interference-plus-noise ratio (SINR) \cite{J:Iterative_Jafar, J:Heath_cooperative} and \textit{v}) maximization of the sum-rate performance \cite{J:Sung_TWC2010, C:Heath_Globecom, C:NegroSlock2011}. 

Although IA attains the optimum sum-rate scaling at the high interference regime, there are certain combinations of SINR levels and channel conditions where it does not \cite{J:Gou_Jafar_LowInterference, J:Veeravalli_LowInterference, C:George_IWSSIP_2012}. For example, the authors in \cite{J:Veeravalli_LowInterference} analyzed conditions for the intended and interference MIMO channels under which treating interference as noise at the receivers is sum-capacity achieving. Very recently, the sum-rate performance results of \cite{C:George_IWSSIP_2012}, for $K$-user MIMO cellular networks with asymmetric average powers and line-of-sight (LOS) conditions among the intended and interference links, demonstrated certain regimes where interference-myopic MIMO transmissions yield superior sum-rate performance to IA. 

Based on the above, it would be desirable to devise a sum-capacity-achieving transmission design for the $K$-user MIMO interference channel that is transparent to the interference conditions. To the best of our knowledge, the majority of the available transmission techniques for the $K$-user interference channel must know a priori the interference conditions so as to choose between the two extremes: treating interference as noise or performing IA. Inspired by the distributed IA algorithm of \cite{J:Iterative_Jafar} and the results of \cite{C:George_IWSSIP_2012}, in this paper we present a distributed iterative algorithm that combines the system-wide minimum MSE (MMSE) criterion with the waterfilling (WF) solution \cite{J:Telatar_Waterfilling} to adjust to the interference levels and channel conditions and maximize accordingly each user's transmission rate. Early numerical evidence corroborates our expectation that the proposed technique reconfigures itself so as to allow the maximum attainable sum rate for the interference levels and channel conditions at hand.

\textit{Notation:} Vectors and matrices are denoted by boldface lowercase letters and boldface capital letters, respectively. The transpose conjugate and the determinant of matrix $\mathbf{A}$ are denoted by $\mathbf{A}^{\rm H}$ and $\det\left(\mathbf{A}\right)$, respectively, whereas $[\mathbf{A}]_{i,j}$ represents the $(i,j)$ element of $\mathbf{A}$ and ${\rm span}(\mathbf{A})$ its column span. $\mathbf{I}_{n}$ is the $n\times n$ identity matrix and ${\rm diag}\{\mathbf{a}\}$ represents a diagonal matrix with vector $\mathbf{a}$ in its main diagonal. In addition, $\mathbf{A}^{(n)}$ represents the $n$th column of $\mathbf{A}$, $||\mathbf{A}||_{\rm F}$ its Frobenius norm, ${\rm Tr}\{\mathbf{A}\}$ its trace and $||\mathbf{a}||$ stands for the Euclidean norm of $\mathbf{a}$. The expectation operator is denoted as $\mathbb{E}\{\cdot\}$ whereas, $X\sim\mathcal{C}\mathcal{N}\left(\mu,\sigma^{2}\right)$ represents a random variable following the complex normal distribution with mean $\mu$ and variance $\sigma^{2}$.

\section{System and Channel Model}\label{sec:SysCha}
We present below the $K$-user MIMO interference system model and the wireless channel model under consideration.

\subsection{System Model}\label{sec:System}
A multiuser MIMO system consisting of $K$ pairs of communicating users is considered. In particular, each transmitting user (Tx) $k$, where $k=1,2,\ldots,K$, equipped with $n_{\rm T}^{[k]}$ antennas wishes to communicate with the $n_{\rm R}^{[k]}$-antenna receiving user (Rx) $k$. All $K$ simultaneous transmissions of symbols $\mathbf{s}_k\in\mathbb{C}^{d_k\times 1}$, with $d_k\leq\min(n_{\rm T}^{[k]}, n_{\rm R}^{[k]})$ $\forall\,k$, are assumed perfectly synchronized and each Tx $k$ processes $\mathbf{s}_k$ with a linear precoding matrix $\mathbf{V}_k\in\mathbb{C}^{n_{\rm T}^{[k]}\times d_k}$ before transmission. In our system model we assume for each $\mathbf{V}_k$ that $||\mathbf{V}_k^{(n)}||=1$ $\forall\,k,n$ with $n=1,2,\ldots,d_k$. For the transmitted power per Tx $k$ it is assumed that $\mathbb{E}\{||\mathbf{V}_k\mathbf{P}_k^{\frac{1}{2}}\mathbf{s}_k||^2\}\leq {\rm P}$ with ${\rm P}$ being the total power constraint per Tx and $\mathbf{P}_k={\rm diag}\{[P_1^{(k)}\,P_2^{(k)}\,\ldots\,P_{d_k}^{(k)}]\}$, where $P_n^{(k)}$ denotes the power allocated to the $n$th data stream at Tx $k$. Without loss of generality, it is assumed throughout this paper that $\mathbb{E}\{\mathbf{s}_k\mathbf{s}_k^{\rm H}\}=\mathbf{I}_{d_k}$ $\forall\,k$. The baseband received signal at Rx $k$ can be mathematically expressed as
\begin{equation}\label{Eq:System}
\mathbf{y}_k = \mathbf{H}_{k,k}\mathbf{V}_k\mathbf{P}_k^{\frac{1}{2}}\mathbf{s}_k + \sum_{\ell=1,\ell\neq k}^K\mathbf{H}_{k,\ell}\mathbf{V}_\ell\mathbf{P}_\ell^{\frac{1}{2}}\mathbf{s}_\ell + \mathbf{n}_k
\end{equation}
where $\mathbf{H}_{k,\ell}\in\mathbb{C}^{n_{\rm R}^{[k]}\times n_{\rm T}^{[\ell]}}$, with $\ell=1,2,\ldots,K$, denotes the channel matrix between Rx $k$ and Tx $\ell$, and $\mathbf{n}_k\in\mathbb{C}^{n_{\rm R}^{[k]}\times 1}$ represents the zero-mean complex additive white Gaussian noise vector with covariance matrix $\sigma^2_k\mathbf{I}_{n_{\rm R}^{[k]}}$. After signal reception, each Rx $k$ is assumed to process $\mathbf{y}_k$ with a linear filter $\mathbf{U}_k\in\mathbb{C}^{n_{\rm R}^{[k]}\times d_k}$ as $\mathbf{U}_k^{\rm H}\mathbf{y}_k$. 

\subsection{Channel Model}\label{sec:Channel}
The flat fading channel model of \cite{C:George_IWSSIP_2012} is assumed for which the channel matrix between Rx $k$ and Tx $\ell$ is given by 
\begin{equation}\label{Eq:Art_Channel}
\mathbf{H}_{k,\ell} =
\left\{
\begin{array}{lr}
\overline{\mathbf{H}}_{k,\ell},&k=\ell\\
\alpha_{k,\ell}\overline{\mathbf{H}}_{k,\ell},&k\neq\ell\\
\end{array}\right.
\end{equation}
where parameter $\alpha_{k,\ell}\in[0,1)$\footnote{For $\alpha_{k,\ell}=1$, \eqref{Eq:Art_Channel} results in the one-branch expression $\mathbf{H}_{k,\ell}=\overline{\mathbf{H}}_{k,\ell}$.} is used for modeling asymme\-tric average powers among the intended and the interference links, and $\overline{\mathbf{H}}_{k,\ell}\in\mathbb{C}^{n_{\rm R}^{[k]}\times n_{\rm T}^{[\ell]}}$, which describes Ricean fading, is defined as
\begin{equation}\label{Eq:Fading}
\overline{\mathbf{H}}_{k,\ell} = \sqrt{\frac{\kappa_{k,\ell}}{\kappa_{k,\ell}+1}}\mathbf{a}_k(\theta_r)\mathbf{a}_\ell(\theta_t)^{\rm H}+\sqrt{\frac{1}{\kappa_{k,\ell}+1}}\overline{\mathbf{H}}_{k,\ell}^{\rm sc}.
\end{equation}
In \eqref{Eq:Fading}, $\kappa_{k,\ell}$ is the Ricean $\kappa$-factor and $\overline{\mathbf{H}}_{k,\ell}^{\rm sc}\in\mathbb{C}^{n_{\rm R}^{[k]}\times n_{\rm T}^{[\ell]}}$ is the scattered component of $\overline{\mathbf{H}}_{k,\ell}$ such that $[\overline{\mathbf{H}}_{k,\ell}^{\rm sc}]_{i,j}\sim\mathcal{CN}(0,1)$ $\forall\,i=1,2,\ldots,n_{\rm R}^{[k]}$ and $\forall\,j=1,2,\ldots,n_{\rm T}^{[\ell]}$. Moreover, $\mathbf{a}_\ell(\theta_t)\in\mathbb{C}^{n_{\rm T}^{[\ell]}\times 1}$ and $\mathbf{a}_k(\theta_r)\in\mathbb{C}^{n_{\rm R}^{[k]}\times 1}$ denote the specular array responses at Tx $\ell$ and Rx $k$, respectively, with $\theta_t$ and $\theta_r$ being the angles of departure and arrival, respectively. 

\section{A Reconfigurable Distributed Algorithm}\label{sec:OurAlgorithm}
In this section we present the motivation and mathematical formulation of the proposed algorithm. A brief discussion on the characteristics of the algorithm is also included in the end. 

\subsection{Motivation}\label{sec:Motivation}
As mentioned earlier, the optimum sum-rate scaling for the $K$-user MIMO interference channel depends on the interfe\-rence levels and channel conditions. Treating interference as noise is preferable at the low interference regime \cite{J:Gou_Jafar_LowInterference, J:Veeravalli_LowInterference, C:George_IWSSIP_2012}, whereas IA achieves the sum capacity at high interference \cite{C:Maddah-Ali, J:Jafar_interference}. To this end, choosing between the latter two strategies requires the a priori knowledge of the interference levels. However, the vast majority of the IA-achieving algorithms requires IA feasibility conditions to be met a priori (see e.g. \cite{C:Maddah-Ali, J:Jafar_interference, C:Schmidt2009, C:Heath_Globecom, J:Iterative_Jafar, J:Sung_TWC2010, J:Heath_cooperative, C:NegroSlock2011, C:Hadi}). For example, to achieve IA for the $3$-user $4\times4$ MIMO interference channel, each Tx must be restricted to send at most $2$ data streams to its intended Rx. On the other hand, \cite{C:George_IWSSIP_2012} demonstrated several low-to-moderate interference scenarios where interference-myopic MIMO transmissions, each aiming at the individual user rate maximization, yield higher sum-rate performance than IA. Finally, a typical feature of the majority of the IA-achieving techniques is the equal power allocation at each Tx's data streams \cite{C:Maddah-Ali, J:Jafar_interference, C:Schmidt2009, C:Heath_Globecom, J:Iterative_Jafar, J:Heath_cooperative, C:NegroSlock2011, C:Hadi}. Inspired by the capacity-achieving strategy for single-user MIMO systems \cite{J:Telatar_Waterfilling} and the sum-rate results for Ricean fading channels presented in \cite{C:George_IWSSIP_2012}, we intuitively expect that the equal power allocation per Tx will be suboptimal in the weak interference regime and under strong LOS conditions.

Motivated by all the above, we present in the following a reconfigurable algorithm (see also \cite{J:Alexandg_IAalg2013}) that implicitly chooses $d_k$ for each Tx $k$ accordingly to the interference levels and channel conditions and jointly designs $\mathbf{V}_{k}$'s, $\mathbf{P}_{k}$'s and $\mathbf{U}_{k}$'s for all transceivers to maximize the sum-rate performance. 

\subsection{Algorithmic Formulation}\label{sec:Algorithm}
The system-wide MSE for the considered $K$-user MIMO interference network is expressed as 
\begin{equation}\label{Eq:J_MMSE}
\mathcal{J}_{\rm MSE} = \sum_{k=1}^K\mathbb{E}\left\{\left\|\mathbf{U}_k^{\rm H}\mathbf{y}_k-\mathbf{s}_k\right\|^2\right\}.
\end{equation}
In this paper we design $\mathbf{V}_k$ and $\mathbf{P}_k$ for each Tx $k$ to maximize each user $k$ rate under the condition that $\mathbf{U}_k$'s  jointly minimize \eqref{Eq:J_MMSE}. In particular, each $\mathbf{V}_k$ is obtained as   
\begin{equation}\label{Eq:Our_Precoding}
\mathbf{V}_k=\mathbf{G}_k\mathbf{F}_k
\end{equation}
with $\mathbf{G}_{k}\in\mathbb{C}^{n_{\rm T}^{[k]}\times d_k}$ and $\mathbf{F}_k\in\mathbb{C}^{d_k\times d_k}$. Given the $\mathbf{U}_k$'s minimizing the system-wide MSE, each $\mathbf{G}_{k}$ is derived as $\mathbf{G}_{k}^{(n)}=\mathbf{E}_{k}^{(n)}/||\mathbf{E}_{k}^{(n)}||$ with $\mathbf{E}_{k}\in\mathbb{C}^{n_{\rm T}^{[k]}\times d_k}$ obtained from 
\begin{equation}\label{Eq:MMSE}
\min_{\{\mathbf{E}_{k}\}_{k=1}^K}\mathcal{J}_{\rm MSE}\,\,{\rm s.\,t.\,}\,\,{\rm Tr}\{\mathbf{E}_k^{\rm H}\mathbf{E}_k\}\leq{\rm P}.
\end{equation} 
Note that for the $\mathcal{J}_{\rm MSE}$ in \eqref{Eq:MMSE}, \eqref{Eq:System} with $\mathbf{E}_{k}=\mathbf{V}_k\mathbf{P}_k^{\frac{1}{2}}$ $\forall\,k$ was utilized. Then, for each Tx $k$, $\mathbf{F}_k$ and $\mathbf{P}_k$ are obtained from
\begin{equation}\label{Eq:maxRate}
\max_{\mathbf{F}_{k},\mathbf{P}_{k}}\mathcal{R}_k\,\,{\rm s.\,t.\,}\,\,\mathbf{F}_k\mathbf{P}_k\mathbf{F}_k^{\rm H}\succeq0\,{\rm and}\,{\rm Tr}\{\mathbf{F}_k\mathbf{P}_k\mathbf{F}_k^{\rm H}\}\leq{\rm P} 
\end{equation}
where $\mathcal{R}_k$ is the instantaneous rate at user $k$ given by
\begin{equation}\label{Eq:SumRate}
\mathcal{R}_k = \log_2\left[\det\left(\mathbf{I}_{n_{\rm R}^{[k]}}+\mathbf{H}_{k,k}\mathbf{G}_k\mathbf{F}_k\mathbf{P}_k\mathbf{F}_k^{\rm H}\mathbf{G}_k^{\rm H}\mathbf{H}_{k,k}^{\rm H}\mathbf{Q}_k^{-1}\right)\right]
\end{equation}
with $\mathbf{Q}_k\in\mathbb{C}^{n_{\rm R}^{[k]}\times n_{\rm R}^{[k]}}$ denoting the interference plus noise covariance matrix at Rx $k$, which is obtained as
\begin{equation}\label{Eq:IpNCov}
\mathbf{Q}_k = \sum_{\ell=1,\ell\neq k}^{K}\mathbf{H}_{k,\ell}\mathbf{G}_{\ell}\mathbf{F}_{\ell}\mathbf{P}_\ell\mathbf{F}_{\ell}^{\rm H}\mathbf{G}_{\ell}^{\rm H}\mathbf{H}_{k,\ell}^{\rm H} + \sigma_k^2\mathbf{I}_{n_{\rm R}^{[k]}}.
\end{equation}

Similar to \cite{J:Iterative_Jafar} we assume in the following reciprocal forward and reverse networks and present a distributed iterative algorithm for obtaining $\mathbf{V}_{k}$'s and $\mathbf{P}_{k}$'s satisfying \eqref{Eq:maxRate} with $\mathbf{U}_{k}$'s minimizing \eqref{Eq:J_MMSE}.

\subsubsection{Forward Network}
In the original forward network, given $\mathbf{V}_{k}$ and $\mathbf{P}_{k}$ $\forall\,k=1,2\ldots,K$, the $n$th column of $\mathbf{U}_k$ at Rx $k$ minimizing \eqref{Eq:J_MMSE} is obtained as
\begin{equation}\label{Eq:Umy}
\mathbf{U}_{k}^{(n)} = \frac{\mathbf{B}_k^{-1}\mathbf{H}_{k,k}\mathbf{V}_{k}^{(n)}}{ \left\|\mathbf{B}_k^{-1}\mathbf{H}_{k,k}\mathbf{V}_{k}^{(n)}\right\|}
\end{equation}
where $\mathbf{B}_k\in\mathbb{C}^{n_{\rm R}^{[k]}\times n_{\rm R}^{[k]}}$ is given by
\begin{equation}\label{Eq:Bmatrix}
\mathbf{B}_k= \sum_{\ell=1}^{K}\mathbf{H}_{k,\ell}\mathbf{V}_{\ell}\mathbf{P}_\ell\mathbf{V}_{\ell}^{\rm H}\mathbf{H}_{k,\ell}^{\rm H} + \sigma^2_k\mathbf{I}_{n_{\rm R}^{[k]}}.
\end{equation}

\subsubsection{Reciprocal Network}
In the reciprocal network, each Rx $k$ utilizes $\mathbf{U}_{k}$ to transmit to its intended Tx $k$. Then, each Tx $k$ computes $\mathbf{V}_{k}$ and $\mathbf{P}_{k}$ through the following two-step procedure. 

\textit{Step I:} Given $\mathbf{U}_{k}$ $\forall\,k=1,2\ldots,K$, each Tx $k$ in the reciprocal network computes its receive filter $\mathbf{E}_{k}$ according to \eqref{Eq:MMSE}. Hence, the $n$th column of $\mathbf{G}_k$ is derived as
\begin{equation}\label{Eq:Emy}
\mathbf{G}_{k}^{(n)} = \frac{\left(\stackrel{\leftarrow}{\mathbf{B}}_k\right)^{-1}\mathbf{H}_{k,k}^{\rm H} \mathbf{U}_{k}^{(n)}}{ \left\|\left(\stackrel{\leftarrow}{\mathbf{B}}_k\right)^{-1}\mathbf{H}_{k,k}^{\rm H} \mathbf{U}_{k}^{(n)}\right\|}
\end{equation}
where $\stackrel{\leftarrow}{\mathbf{B}}_k\in\mathbb{C}^{n_{\rm T}^{[k]}\times n_{\rm T}^{[k]}}$ is given by
\begin{equation}\label{Eq:B_Rev}
\stackrel{\leftarrow}{\mathbf{B}}_k = \sum_{\ell=1}^{K}\frac{\rm P}{d_\ell}\mathbf{H}_{\ell,k}^{\rm H}\mathbf{U}_{\ell}\mathbf{U}_{\ell}^{\rm H}\mathbf{H}_{\ell,k} + \mu_k\mathbf{I}_{n_{\rm T}^{[k]}}.
\end{equation}
In \eqref{Eq:B_Rev}, parameter $\mu_k$ is calculated so that the power constraint at Tx $k$ is satisfied \cite{J:Heath_cooperative}.

\textit{Step II:} After obtaining all $\mathbf{G}_{k}$'s from \eqref{Eq:Emy} and given $\mathbf{P}_{\ell}$ $\forall\,\ell\neq k$, each Tx $k$ computes its optimum precoding and power allocation matrices for its effective channel $\mathbf{H}_{k,k}\mathbf{G}_{k}$ assuming knowledge of $\mathbf{Q}_k$, which is given by \eqref{Eq:IpNCov}. In particular, the singular value decomposition (SVD) of $\mathbf{H}_{k,k}\mathbf{G}_{k}$ after noise prewhitening is derived as
\begin{equation}\label{Eq:SVD_Prewhitened}
\mathbf{Q}_k^{-\frac{1}{2}}\mathbf{H}_{k,k}\mathbf{G}_{k} = \mathbf{W}_k\mathbf{\Lambda}_k\mathbf{F}_k^{\rm H}
\end{equation}
where $\mathbf{W}_k\in\mathbb{C}^{n_{\rm R}^{[k]}\times n_{\rm R}^{[k]}}$ and $\mathbf{\Lambda}_k\in\mathbb{C}^{n_{\rm R}^{[k]}\times d_k}$, and $\mathbf{F}_k$ is the optimum precoding matrix for $\mathbf{H}_{k,k}\mathbf{G}_{k}$ to be utilized in \eqref{Eq:Our_Precoding}. The power allocation $\mathbf{P}_k$ for each Tx $k$ data streams is finally derived from the WF solution for the channel $\mathbf{Q}_k^{-\frac{1}{2}}\mathbf{H}_{k,k}\mathbf{G}_{k}$.

The proposed reconfigurable distributed iterative algorithm is summarized in Algorithm~\ref{MyAlgorithm}.

\begin{algorithm}[!t]
\caption{Reconfigurable Precoding}\label{MyAlgorithm}
\begin{algorithmic}[1]
\STATE \textbf{initialization:} Set $d_k=\min(n_{\rm T}^{[k]}, n_{\rm R}^{[k]})$ $\forall\,k=1,2,\ldots,K$ and start with arbitrary unit-column $\mathbf{V}_k\in\mathbb{C}^{n_{\rm T}^{[k]}\times d_k}$ and $\mathbf{P}_k={\rm P}/d_k \mathbf{I}_{n_{\rm T}^{[k]}}$ 
\STATE Begin iteration\\
\textbf{\textit{Forward Network}}
\STATE Compute $\mathbf{B}_k$ at each Rx $k$ according to \eqref{Eq:Bmatrix}
\STATE Obtain each Rx $k$ MMSE filter $\mathbf{U}_k$ using \eqref{Eq:Umy}\\
\textbf{\textit{Reciprocal Network}}\\
\textit{Step I:}
\STATE Compute $\stackrel{\leftarrow}{\mathbf{B}}_k$ at each Tx $k$ according to \eqref{Eq:B_Rev}
\STATE Obtain each Tx $k$ MMSE-based $\mathbf{G}_k$'s using \eqref{Eq:Emy}\\
\textit{Step II:}
\STATE Compute $\mathbf{Q}_k$ at each Rx $k$ according to \eqref{Eq:IpNCov}
\STATE Perform SVD to each $\mathbf{Q}_k^{-\frac{1}{2}}\mathbf{H}_{k,k}\mathbf{G}_{k}$ according to \eqref{Eq:SVD_Prewhitened}
\STATE Obtain each Tx $k$ precoding matrix $\mathbf{V}_k$ using \eqref{Eq:Our_Precoding}
\STATE Compute $\mathbf{P}_k$ for each Tx $k$ from the WF solution for the effective channel $\mathbf{Q}_k^{-\frac{1}{2}}\mathbf{H}_{k,k}\mathbf{G}_{k}$ 
\STATE Repeat until sum rate converges, or until the number of iterations reaches a predefined limit
\end{algorithmic}
\end{algorithm}

\subsection{Discussion}\label{sec:Convergence}
The proposed algorithm capitalizes on the reciprocity of wireless channels, such as when time-division duplexing communication is used, to design, in a distributed manner, $\mathbf{V}_{k}$'s and $\mathbf{P}_{k}$'s maximizing the individual user rates as well as $\mathbf{U}_{k}$'s minimizing the system-wide MSE. Similar to \cite{J:Iterative_Jafar}, in the forward network each Rx $k$ obtains $\mathbf{U}_{k}$ using only local information, i$.$e$.$ $\mathbf{H}_{k,k}\mathbf{V}_{k}$ and $\mathbf{B}_{k}$. For the reciprocal network, in \textit{Step I}, each Tx $k$ computes its $\mathbf{G}_{k}$ using the locally available $\mathbf{H}_{k,k}^{\rm H} \mathbf{U}_{k}$ and $\stackrel{\leftarrow}{\mathbf{B}}_{k}$. Then, in \textit{Step II}, each Tx $k$ utilizes $\mathbf{Q}_{k}$ available at its intended Rx $k$ to obtain $\mathbf{F}_{k}$ and $\mathbf{P}_{k}$ maximizing its own rate $\mathcal{R}_k$. More specifically, \textit{Step I} computes subspaces ${\rm span}(\mathbf{H}_{k,k}\mathbf{G}_{k})$	$\forall\,k$ whereas, \textit{Step II} obtains the WF solution for each effective channel $\mathbf{Q}_k^{-\frac{1}{2}}\mathbf{H}_{k,k}\mathbf{G}_{k}$. For networks with low average power interference links and as it will be shown later on, it is sufficient to use \textit{Step II} of Algorithm~\ref{MyAlgorithm} without $\mathbf{Q}_{k}$. This variation of Algorithm~\ref{MyAlgorithm}, termed as Reconfigurable Myopic Precoding (Algorithm~2), does not utilize statement $7$ in Algorithm~\ref{MyAlgorithm} and uses statements $8$ and $10$ without $\mathbf{Q}_{k}$. Obviously, Algorithm~2 does not need each Rx $k$ to feedback $\mathbf{Q}_{k}$ at its Tx $k$ and hence the algorithmic complexity is similar to the algorithms presented in \cite{J:Iterative_Jafar}.

The adjustment of Algorithm~\ref{MyAlgorithm} to the interference levels and channel conditions lies on the WF solution utilized in \textit{Step II}. In particular, the algorithm is initialized with the maximum allowable number of data streams per Tx $k$, i$.$e$.$ $d_k=\min(n_{\rm T}^{[k]}, n_{\rm R}^{[k]})$. In each iteration the receive filters minimizing the system-wide MMSE are obtained in the forward network and, in the reciprocal network, the WF solution provides the precoding and power allocation matrices for each Tx $k$, thus implicitly $d_k$, that maximize $\mathcal{R}_k$. Our numerous computer simulation results indicated that the instantaneous sum-rate performance of our algorithm, obtained using \eqref{Eq:SumRate} as $\sum_{k=1}^K\mathcal{R}_k$, converges often to a maximum value.
\begin{figure}[!t]
\centering
\includegraphics[height=2.8in,width=2.8in]{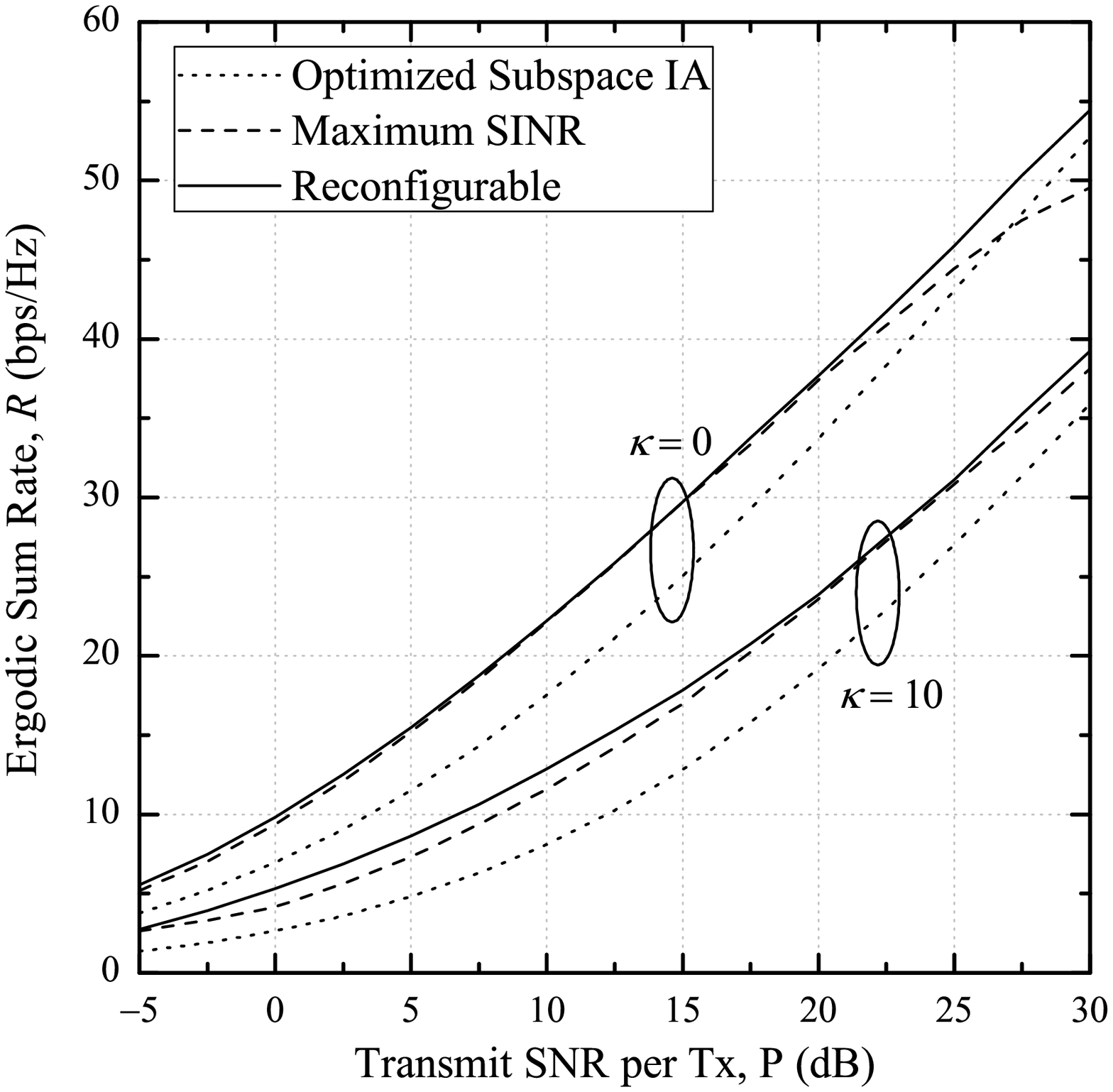}
\caption{Sum-rate performance, $R$, versus transmit SNR per Tx, ${\rm P}$, for $3$-user $4\times4$ MIMO networks over Rayleigh and Ricean fading with $\alpha=1$.}
\label{Fig:SRa1Ray}
\end{figure} 

\section{Sum-Rate Performance Results}\label{sec:Performance}
This section presents numerical simulations for the sum-rate performance of the proposed reconfigurable distributed iterative algorithm for the $3$-user $4\times4$ MIMO interference channel. For comparison purposes, sum-rate computer simulations for the algorithms: \textit{i}) optimized subspace IA \cite[Sec. IV.B.1]{J:Sung_TWC2010} and \textit{ii}) distributed per stream SINR maximization (Maximum SINR) \cite{J:Jafar_interference} are also shown. In particular, we have simulated the ergodic sum-rate performance defined as \cite{J:Heath_cooperative}
\begin{equation}\label{Eq:Erg_SumRate}
R = \mathbb{E}\left\{\sum_{k=1}^K\log_2\left[\det\left(\mathbf{I}_{n_{\rm R}^{[k]}}+\mathbf{H}_{k,k}\mathbf{V}_k\mathbf{P}_k\mathbf{V}_k^{\rm H}\mathbf{H}_{k,k}^{\rm H}\mathbf{Q}_k^{-1}\right)\right]\right\}.
\end{equation}
The averaging in \eqref{Eq:Erg_SumRate} was evaluated via Monte Carlo simulations for 100 independent channel realizations and the channels were normalized as $\mathbb{E}\left\{\left\|\mathbf{H}_{k,j}\right\|_{\rm F}^2\right\} = n_{\rm R}^{[k]}n_{\rm T}^{[j]}$ $\forall\,k,j=1,2,\ldots,K$. Without loss of generality, for the channel model in Sec~\ref{sec:Channel}, we have assumed that $\alpha_{k,\ell}=\alpha$ and $\kappa_{k,\ell}=\kappa$ $\forall\,k,\ell$. For all algorithms, $\mathbf{V}_k$'s and $\mathbf{U}_k$'s were randomly initialized with unit norm columns. The proposed algorithm was also initialized with $\mathbf{P}_k={\rm P}/4 \mathbf{I}_{4}$ $\forall\,k=1,2$ and $3$ whereas, for the optimized subspace IA and Maximum SINR, IA feasibility conditions were set a priori, i$.$e$.$ $d_k=2$ $\forall\,k=1,2$ and $3$. In addition, for scenarios with low average power interference links, we have simulated the ergodic sum rate of a genie-aided Maximum SINR that utilizes $d_k=4$ $\forall\,k=1,2$ and $3$. A maximum of $1000$ iterations was used per distributed iterative algorithm and each algorithm was declared converged when the difference in its objective function between two successive iterations was less than $10^{-4}$.

\begin{figure}
\centering
\includegraphics[height=2.8in,width=2.8in]{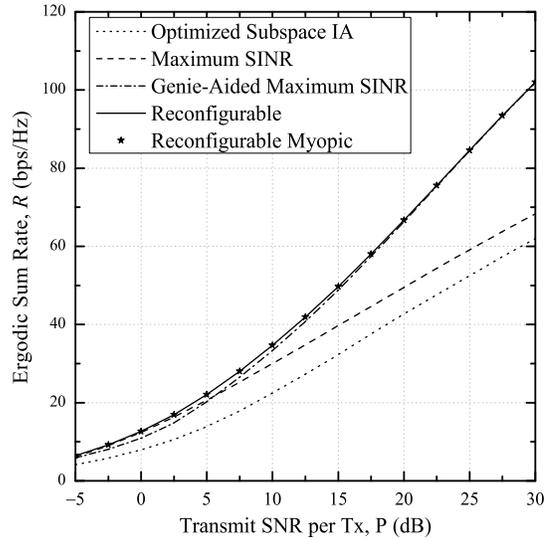}
\caption{Sum-rate performance, $R$, versus transmit SNR per Tx, ${\rm P}$, for $3$-user $4\times4$ MIMO networks over Rayleigh fading with $\alpha=10^{-2}$.}
\label{Fig:SRa10-3Ray}
\end{figure} 
As shown in Fig.~\ref{Fig:SRa1Ray} for various Ricean fading channels with $\alpha=1$, the sum rate of the reconfigurable Algorithm~\ref{MyAlgorithm} is similar to that of the Maximum SINR. In particular, for low signal-to-noise ratio (SNR) values, the proposed algorithm computes Tx filters that maximize signal powers over the noise power. The latter maximizations include optimum power allocation, resulting in a slightly higher $R$ than that of the Maximum SINR in the noise-limited regime. For Rayleigh fading this behavior happens for SNR values ranging from $-5$ to $5$ dB whereas, for Ricean fading channels with $\kappa=10$, the gains from power allocation are higher and for a wider range of the SNR. To this end, optimum power allocation seems to result in higher $R$ gains as LOS conditions among all users in the network become stronger. More importantly, in the interference-limited regime, the proposed algorithm adjusts itself so as to achieve IA, which is rate-scaling optimal under strong interference conditions. 

\begin{figure}[!t]
\centering
\includegraphics[height=2.8in,width=2.8in]{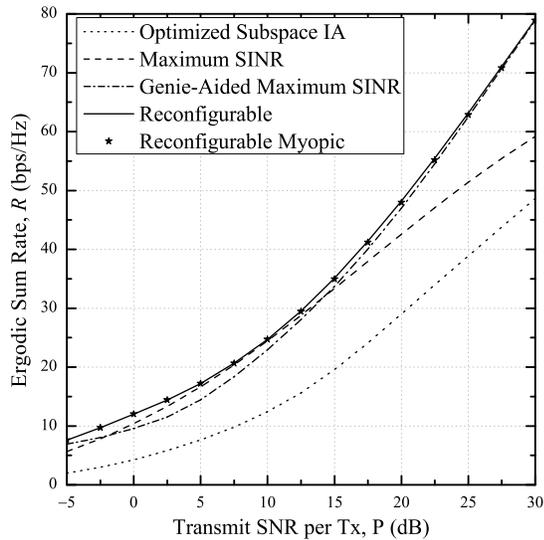}
\caption{Sum-rate performance, $R$, versus transmit SNR per Tx, ${\rm P}$, for $3$-user $4\times4$ MIMO networks over Ricean fading with $\kappa=10$ and $\alpha=10^{-2}$.}
\label{Fig:SRa10-3Rice10}
\end{figure}

\begin{figure}[!t]
\centering
\includegraphics[height=2.8in,width=2.8in]{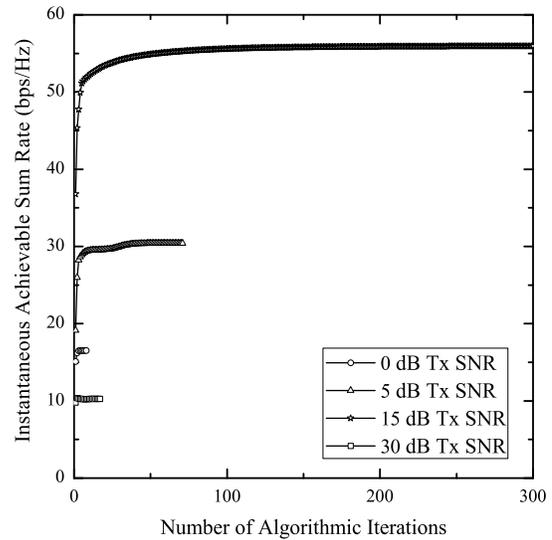}
\caption{Convergence of the instantaneous achievable sum rate for $3$-user $4\times4$ MIMO networks over Rayleigh fading with $\alpha=1$.}
\label{Fig:CONVERGENCEa1_Ray}
\end{figure}

Figures~\ref{Fig:SRa10-3Ray} and \ref{Fig:SRa10-3Rice10} depict the sum rate versus SNR for Ricean fading channels with $\kappa=0$ and $10$, respectively, and for low average power interference links with $\alpha=10^{-2}$. As shown, the reconfigurable Algorithm~\ref{MyAlgorithm} adjusts itself to this interference scenario as well as to the channel conditions, and achieves a sum-rate scaling of $3$ at high SNRs. More importantly, Algorithm~\ref{MyAlgorithm} outperforms both the Maximum SINR and genie-aided Maximum SINR at low-to-moderate SNRs. As expected, the Maximum SINR, which is restricted to the IA feasibility conditions, achieves only a sum-rate scaling of $1.5$ at high SNRs whereas, the genie-aided Maximum SINR results in poor performance at low SNRs due to the equal power allocation. Within these figures, the performance of the reconfigurable myopic Algorithm~$2$ is also illustrated and it is shown that it yields similar sum rate to Algorithm~\ref{MyAlgorithm}. Clearly, for interference levels with $\alpha=10^{-2}$, the low complexity Algorithm~$2$ might be used instead of Algorithm~\ref{MyAlgorithm}.
Finally, Fig.~\ref{Fig:CONVERGENCEa1_Ray} depicts the convergence of the instantaneous achievable sum rate for Algorithm~\ref{MyAlgorithm} over Rayleigh fading channels with $\alpha=1$ and for various values of the SNR. As shown, the sum rate converges fast to a maximum value and the speed of convergence depends on the SNR. In particular, as the SNR increases, more algorithmic iterations are needed for the convergence of Algorithm~\ref{MyAlgorithm}. 

\section{Conclusions}\label{sec:Conclusion}
In this paper, a novel reconfigurable distributed iterative algorithm for $K$-user MIMO interference networks is presented. The proposed algorithm combines the system-wide MMSE with the WF solution to adjust to the interference levels and channel conditions and maximize accordingly each user's transmission rate. As shown, in the interference-limited regime, our algorithm adjusts itself so as to achieve the IA scaling whereas, in the low-to-moderate interference regime, it chooses interference-myopic MIMO transmissions. Furthermore, for all investigated interference cases and channel conditions, it was shown that the sum rate of the proposed algorithm is higher than that of all other considered algorithms.

\section*{Acknowledgment}
This work has been supported by the European Union Future and Emerging Technologies (FET) Project HiATUS. The project HiATUS acknowledges the financial support of the FET programme, within the Seventh Framework Programme for Research of the European Commission, under FET-open grant number 265578.

\bibliographystyle{IEEEtran}
\bibliography{IEEEabrv,refs}

\begin{thebibliography}{10}
\providecommand{\url}[1]{#1}
\csname url@samestyle\endcsname
\providecommand{\newblock}{\relax}
\providecommand{\bibinfo}[2]{#2}
\providecommand{\BIBentrySTDinterwordspacing}{\spaceskip=0pt\relax}
\providecommand{\BIBentryALTinterwordstretchfactor}{4}
\providecommand{\BIBentryALTinterwordspacing}{\spaceskip=\fontdimen2\font plus
\BIBentryALTinterwordstretchfactor\fontdimen3\font minus
  \fontdimen4\font\relax}
\providecommand{\BIBforeignlanguage}[2]{{%
\expandafter\ifx\csname l@#1\endcsname\relax
\typeout{** WARNING: IEEEtran.bst: No hyphenation pattern has been}%
\typeout{** loaded for the language `#1'. Using the pattern for}%
\typeout{** the default language instead.}%
\else
\language=\csname l@#1\endcsname
\fi
#2}}
\providecommand{\BIBdecl}{\relax}
\BIBdecl

\bibitem{C:Maddah-Ali}
M.~Maddah-Ali, A.~Motahari, and A.~Khandani, ``Signaling over {MIMO} multi-base
  systems: {C}ombination of multi-access and broadcast schemes,'' in
  \emph{Proc. IEEE ISIT}, Seattle, USA, 9-14 Jul. 2006, pp. 2104–--2108.

\bibitem{J:Jafar_interference}
V.~Cadambe and S.~A. Jafar, ``Interference alignment and degrees of freedom of
  the {$K$}-user interference channel,'' \emph{IEEE Trans. on Inf. Theory},
  vol.~54, no.~8, pp. 3425–--3441, Aug. 2008.

\bibitem{C:Schmidt2009}
D.~A. Schmidt, C.~Shiy, R.~A. Berryy, M.~L. Honigy, and W.~Utschick, ``Minimum
  mean squared error interference alignment,'' in \emph{Proc. Asilomar CSSC},
  Pasific Grove, USA, 1-4 Nov. 2009, pp. 1106--1110.

\bibitem{C:Heath_Globecom}
I.~Santamaria, O.~Gonzalez, R.~W. Heath, Jr., and S.~W. Peters, ``Maximum
  sum-rate interference alignment algorithms for {MIMO} channels,'' in
  \emph{Proc. IEEE GLOBECOM}, Miami, USA, 6-10 Dec. 2010.

\bibitem{J:Iterative_Jafar}
K.~Gomadam, V.~R. Cadambe, and S.~A. Jafar, ``Distributed numerical approach to
  interference alignment and applications to wireless interference networks,''
  \emph{IEEE Trans. on Inf. Theory}, vol.~57, no.~6, pp. 3309–--3322, Jun.
  2011.

\bibitem{J:Sung_TWC2010}
H.~Sung, S.-H. Park, K.-J. Lee, and I.~Lee, ``Linear precoder designs for
  {$K$}-user interference channels,'' \emph{IEEE Trans. on Wireless Commun.},
  vol.~9, no.~1, pp. 291--301, Jan. 2010.

\bibitem{J:Heath_cooperative}
S.~W. Peters and R.~W. Heath, Jr., ``Cooperative algorithms for {MIMO}
  interference channels,'' \emph{IEEE Trans. on Veh. Tech.}, vol.~60, no.~1,
  pp. 206--218, Jan. 2011.

\bibitem{C:NegroSlock2011}
F.~Negro, I.~Ghauri, and D.~T.~M. Slock, ``Deterministic annealing design and
  analysis of the noisy {MIMO} interference channel,'' in \emph{Proc. IEEE
  ITA}, San Diego, USA, 6-11 Feb. 2011.

\bibitem{C:Hadi}
H.~G. Ghauch and C.~B. Papadias, ``Interference alignment: {A} one-sided
  approach,'' in \emph{Proc. IEEE GLOBECOM}, Houston, USA, 5-9 Dec. 2011.

\bibitem{J:Luo_TIT_2012}
M.~Razaviyayn, M.~Sanjabi, and Z.-Q. Luo, ``Linear transceiver design for
  interference alignment: {C}omplexity and computation,'' \emph{IEEE Trans. on
  Inf. Theory}, vol.~58, no.~5, pp. 2896--2910, May 2012.

\bibitem{J:Yetis}
C.~M. Yetis, T.~Gou, S.~A. Jafar, and A.~H. Kayran, ``On feasibility of
  interference alignment in {MIMO} interference networks,'' \emph{IEEE Trans.
  on Signal Process.}, vol.~58, no.~9, pp. 4771--4782, Sep. 2010.

\bibitem{J:Gou_KMNchannels}
T.~Gou and S.~A. Jafar, ``Degrees of freedom of the {$K$}-user {$M \times N$}
  {MIMO} interference channel,'' \emph{IEEE Trans. on Inf. Theory}, vol.~56,
  no.~12, pp. 6040–--6057, Dec. 2010.

\bibitem{C:Heath_Minimization}
S.~W. Peters and R.~W. Heath, Jr., ``Interference alignment via alternating
  minimization,'' in \emph{Proc. IEEE ICASSP}, Taiwan, Taipei, 19-24 Apr. 2009,
  pp. 2445--2448.

\bibitem{J:Gou_Jafar_LowInterference}
T.~Gou and S.~A. Jafar, ``Sum capacity of a class of symmetric {SIMO}
  {G}aussian interference channels within $\mathcal{O}(1)$,'' \emph{IEEE Trans.
  on Inf. Theory}, vol.~57, no.~4, pp. 1932--1958, Apr. 2011.

\bibitem{J:Veeravalli_LowInterference}
V.~S. Annapureddy and V.~V. Veeravalli, ``Sum capacity of {MIMO} interference
  channels in the low interference regime,'' \emph{IEEE Trans. on Inf. Theory},
  vol.~57, no.~5, pp. 2565--2581, May 2011.

\bibitem{C:George_IWSSIP_2012}
G.~C. Alexandropoulos, S.~Papaharalabos, and C.~B. Papadias, ``On the
  performance of interference alignment under weak interference conditions,''
  in \emph{Proc. IEEE IWSSIP}, Vienna, Austria, 11-13 Apr. 2012.

\bibitem{J:Telatar_Waterfilling}
E.~Telatar, ``Capacity of multi-antenna {G}aussian channels,'' \emph{European
  Trans. on Telecommun.}, vol.~10, no.~6, pp. 585--595, Dec. 1999.

\bibitem{J:Alexandg_IAalg2013}
G.~C. Alexandropoulos and C.~B. Papadias, ``A reconfigurable iterative
  algorithm for the {$K$}-user {MIMO} interference channel,'' \emph{Signal
  Processing (Elsevier)}, under revision, 2013.

\end{thebibliography}

\end{document}